\documentstyle[12pt]{article}

\newcommand{\nc}{\newcommand}
\nc{\beq}{\begin{equation}}
\nc{\eeq}{\end{equation}}
\nc{\beqa}{\begin{eqnarray}}
\nc{\eeqa}{\end{eqnarray}}

\input epsf
\newwrite\ffile\global\newcount\figno \global\figno=1
\def\writedef#1{}

\def\figin{\epsfcheck\figin}\def\figins{\epsfcheck\figins}
\def\epsfcheck{\ifx\epsfbox\UnDeFiNeD
\message{(NO epsf.tex, FIGURES WILL BE IGNORED)}

\gdef\figin##1{\vskip2in}\gdef\figins##1{\hskip.5in}% blank space instead
\else\message{(FIGURES WILL BE INCLUDED)}%
\gdef\figin##1{##1}\gdef\figins##1{##1}\fi}

\def\figinsert{}
\def\ifig#1#2#3{\xdef#1{fig.~\the\figno}
\writedef{#1\leftbracket fig.\noexpand~\the\figno}%
\figinsert\figin{\centerline{#3}}\medskip\centerline{\vbox{\baselineskip12pt
\advance\hsize by -1truein\center\footnotesize{  Fig.~\the\figno.} #2}}
\bigskip\endinsert\global\advance\figno by1}
\def\endinsert{}

\begin{document}

\title{\large{\bf  A note on bosonic open strings in constant
 $B$ field}}

\author{Ansar Fayyazuddin\thanks{ansar@physto.se}\,\,\,\,\,\,\,
 and \,\,\,\,\,
Maxim Zabzine\thanks{zabzin@physto.se}\\
\\
{\small{Institute of Theoretical Physics, University of Stockholm,}}\\
{\small{ Box 6730, S-11385 Stockholm, Sweden}}\\}

\date{}

%\date{September 1997}

\maketitle

\begin{picture}(0,0)(0,0)
\put(390,245){USITP-99-8}
\put(390,225){hep-th/9911018}
%\put(390,225){draft}
\end{picture}
\vspace{-24pt}

\begin{abstract}
 We sketch the main steps of old covariant quantization of bosonic
 open strings in a constant $B$ field background. 
 We comment on its space-time symmetries and the induced 
 effective metric.
% The model has the full group of
% translations but the broken Lorentz group. We discuss the question of
% invariant metric in the model. 
 The low-energy spectrum is evaluated and the appearance of a new
 non-commutative gauge symmetry is addressed. 
 %As well as symmetries of the massless
 %spectrum is considered.
 %Some problems which arise within
 %this quantization are briefly discussed. 
\end{abstract}

\date{}

\maketitle

\newpage

\section{Introduction}

Starting with \cite{Connes:1998cr} evidence has been 
gathering 
that ideas of noncommutative geometry 
 may play an important role in string theory (for a review
 and more references see \cite{Seiberg:1999vs}). Especially interesting is
 the model of open strings propogating in a constant two-form 
 (${\cal B}$-field) background. This model attracted attention since 
it can be studied from a perturbative string point 
of view.  Previous studies show that this model is related to 
noncommutativity of D-branes \cite{Chu:1998qz}, \cite{Ardalan:1998ce}
  and in the zero slope limit to
 noncommutative Yang-Mills theory \cite{Seiberg:1999vs}. 
 In this note we would like to study
 this system  from the old perturbative string perspective concentrating
 on its symmetries.        

 In the conformal gauge the model has the following action
\beq
\label{a0}
 S = - \frac{1}{4\pi\alpha'} \int\limits_{\Sigma}
  d^2\sigma\,[ \eta_{\mu\nu} \partial_\alpha X^\mu
 \partial^{\alpha} X^\nu - \epsilon^{\alpha\beta} {\cal B}_{\mu\nu} 
 \partial_\alpha X^\mu \partial_\beta X^\nu ]
 + \int\limits_{\partial \Sigma} d\tau A_\mu \dot{X}^\mu
\eeq
 where $\Sigma$ is an oriented world-sheet with boundaries and signature
 $(-1,1)$. For the open
 string case the ${\cal B}$-field should appear in a gauge 
 invariant combination with the $U(1)$ gauge field $B_{\mu\nu} = 
 {\cal B} + 2\pi \alpha' \partial_{[\mu} A_{\nu]}$. Thus the action  
 has the form 
\beq
\label{a1}
S = - \frac{1}{4\pi\alpha'} \int d^2\sigma\,[\eta_{\mu\nu} 
 \partial_\alpha X^\mu
 \partial^{\alpha} X^\nu - \epsilon^{\alpha\beta} B_{\mu\nu} 
 \partial_\alpha X^\mu \partial_\beta X^\nu ]
\eeq
  where $B$ is assumed to be constant. 
 The equation of motion and the boundary conditions are given by
 the following expressions
\beq
\label{a2}
\partial_\alpha \partial^\alpha X^\mu=0,\,\,\,\,\,\,\,\,\,\,\,\,\,\,\,
\eta_{\mu\nu} X'^\nu + B_{\mu\nu} \dot{X}^\nu |_{\sigma=0,\pi} =0.
\eeq
 In what follows we will not impose any pure Dirichlet boundary 
 conditions. All results can be strightforwardly generalized to 
  the case with pure Dirichlet conditions. Thus using modern
  language one can say that we are looking at a D25-brane with constant
  $B$-field. 

 The paper is organized as follows: In section 2 we sketch some steps of
 the 
 covariant quantization of the model. In sections 3 and 4 the symmetries 
 and the mass spectrum are discussed. In the last two sections we  
 consider noncommutative gauge transformations of massless states 
 and explain how it might be related to deformation quantization.

\section{The old covariant quantization}

 In this section we sketch some steps of the old covariant approach
 to quantization. Essential parts of this calculations have already 
 appeared in
 \cite{Chu:1998qz}. However, for the sake of completness we prefer to 
 go through this calculations once more. We will write the equations in
 such a way that one can analyze the possible modifications of them
 in different situations. 

The general form of the bulk equation of motion 
 $\partial_\alpha\partial^\alpha X^\mu=0$ is given by
\beq
\label{a3}
X^\mu(\tau,\sigma) = q^\mu + a_0^\mu \tau + b_0^\mu \sigma +
 \sum\limits_{n\neq 0} \frac{e^{-in\tau}}{n} ( i a_n^\mu \cos n\sigma
 + b_n^\mu \sin n\sigma)
\eeq
 By imposing boundary conditions (\ref{a2})  on (\ref{a3}) 
 we get the following solution
\beq
\label{a4}
X^\mu(\tau,\sigma) = q^\mu + (a_0^\mu \tau - B^{\mu}_{\,\,\nu}
  a_0^\nu \sigma)  +
 \sum\limits_{n\neq 0} \frac{e^{-in\tau}}{n} ( i a_n^\mu \cos n\sigma
  -  B^{\mu}_{\,\,\nu}
  a_n^\nu \sin n\sigma)
\eeq
 One can check that the form of Virasoro conditions are not 
 modified
\beq
\label{a5}
\frac{1}{4\alpha'}(\dot{X} \pm X')^2 = 0
\eeq
 and using the expansion (\ref{a4}) the constraints (\ref{a5})
 can be expanded as follows
\beq
\label{a6}
 L_k = \frac{1}{4\alpha'}
 \sum\limits_{n=-\infty}^{\infty} G_{\nu\rho} a_n^\nu a_{k-n}^\rho,
 \,\,\,\,\,\,\,\,\,\,\,\,\,\,\,\,\,\,
 G_{\nu\rho} = \eta_{\nu\rho} - B_{\nu\mu} \eta^{\mu\sigma} B_{\sigma\rho}
\eeq
 where $G_{\mu\rho}$ coincides with the notation of
\footnote{In
  Euclidean signature the matrix $G_{\mu\nu}$ has the same form.} 
 \cite{Seiberg:1999vs}.
  From the action (\ref{a1}) one can define the canonical momentum density
\beq
\label{b1}
 P_{\mu}= \frac{1}{2\pi\alpha'} ( \eta_{\mu\nu} \dot{X}^\nu +
 B_{\mu\nu} X'^\nu ) 
\eeq
 which has the following expansion
\beq
\label{b2}
 P_\mu = \frac{1}{2\pi\alpha'} \sum\limits_{n=-\infty}^{\infty}
 G_{\mu\nu}  a^\nu_n \,
e^{-in\tau} \cos n\sigma  
\eeq
 The canonical commutation relations have the standard form 
$$[P_\mu(\tau,\sigma), X^\nu(\tau, \sigma')] = -i \delta_\mu^\nu 
 \delta(\sigma-\sigma'),$$
\beq
\label{b3}
 [ X^\mu(\tau, \sigma),  X^\nu(\tau, \sigma')] = 
[P_\mu(\tau,\sigma), P_\nu(\tau,\sigma')] = 0.
\eeq
 They imply the commutation relations for the modes.
 From this point we would like to be more careful. 
 One can start from the relations $[P_\mu(\tau,\sigma), 
 P_\nu(\tau,\sigma')] = 0$ and $[P_\mu(\tau,\sigma), 
 X^\nu(\tau, \sigma')] = -i \delta_\mu^\nu 
 \delta(\sigma-\sigma')$. These relations fix uniquely  
 the following commutators 
\beq
\label{b6}
 G_{\mu\rho} [a_n^\rho, a_m^\nu] = 2\alpha' n \delta_\mu^\nu \delta_{m+n},
\,\,\,\,\,\,\,\,\,\,\,\,\,\,\,\,\,\,\,\,\,\,
G_{\mu\rho} [a_n^\rho, q^\nu] = -2i\alpha' \delta_\mu^\nu \delta_n
\eeq
 So far we have not assumed anything definite about the matrix $G_{\mu\rho}$.
 To proceed one should assume that
 $G_{\mu\rho}$ is invertible\footnote{Invertibility is guaranteed
as long as $B_{0i}$ components are not too large.  See below.}. 
 Now one can look at the commutator $ [ X^\mu(\tau, \sigma),  
 X^\nu(\tau, \sigma')] = 0$. Using the mode expansion and
 commutation relations (\ref{b6}) we get
\beq
\label{b8}
[ X^\mu(\tau, \sigma),  X^\nu(\tau, \sigma')] = 
 [q^\mu, q^\nu] - 2i\alpha' \theta^{\mu\nu} (\sigma +\sigma') -
  2i\alpha' \theta^{\mu\nu} \sum\limits_{n\neq 0}\frac{1}{n} 
 \sin n(\sigma+ \sigma') = 0
\eeq
 where $\theta^{\mu\nu} = -B^\mu_{\,\,\sigma} (G^{-1})^{\sigma \nu}$
 and it coincides with the notation of \cite{Seiberg:1999vs}.
 Using the following Fourier expansion on the interval $(0,2\pi)$ for
 $(\sigma+\sigma')$ 
\beq
\label{b9}
\sum\limits_{n\neq 0}\frac{1}{n} \sin n(\sigma+
 \sigma') = \pi - (\sigma+\sigma') 
\eeq
 we can fix commutators for $q^\mu$. 
 Thus by requiring the commutation relations to be valid at all 
 nonboundary points we get the commutators
\beq
\label{b10}
 [a_n^\rho, a_m^\nu] = 2\alpha' n \delta_{n+m} (G^{-1})^{\rho\nu},
\,\,\,\,\,\,\,\,\,\,\,\,\,\,
 [a_n^\rho, q^\nu] = -2i\alpha' \delta_n (G^{-1})^{\rho\nu},
\,\,\,\,\,\,\,\,\,\,\,\,\,\,
 [q^\mu, q^\nu] = 2\pi i\alpha' \theta^{\mu\nu}
\eeq

 Using (\ref{a6}) and (\ref{b10}) one can show that $L_k$
 satisfy the Virasoro algebra
\beq
\label{b11}
 [L_k, L_n] = (k-n) L_{k+n} + A(n) \delta_{k+n}
\eeq
 where using standard arguments \cite{Green:1987sp} 
 one can show that the anomaly
 has the same value as in the usual ($B=0$) case.

\section{Symmetries of the model}

 Now let us look at the symmetries of the action (\ref{a1}). Naively
 one might expect the full ten dimensional Poincar\'e group to be a
symmetry of the action. The action
 (\ref{a1}) is certainly invariant under ten dimensional translations
 $X^\mu \rightarrow X^\mu + b^\mu$. Using the Noether theorem we
 can find the correspondig conserved current. The zero component
 of this current coincides with the momentum density (\ref{b1}). Thus
 one can define the momentum operator for the string as follows
\beq
\label{k1}
 p_\mu = \int\limits_{0}^{\pi} d\sigma\, P_\mu(\tau,\sigma) = 
 \frac{1}{2\alpha'} G_{\mu\nu} a_0^\nu.
\eeq
 One can check the conservation of momentum explicitly
\beq
\label{k2}
 \dot{p}_\mu = \frac{1}{2\pi\alpha'} \int\limits_{0}^{\pi} d\sigma\,
 (\eta_{\mu\nu} \ddot{X}^\nu + B_{\mu\nu} \dot{X}'^\nu) =
 \frac{1}{2\pi\alpha'} (\eta_{\mu\nu} X'^{\nu} + B_{\mu\nu}
 \dot{X}^\nu)|_{\sigma=0,\pi} = 0.
\eeq
 We can rewrite the commutations relations
 for the zero modes as follows
\beq
\label{k3}
 [q^\mu, p_\nu] = i \delta_\nu^\mu,\,\,\,\,\,\,\,\,\,\,\,\,\,\,\,
 [q^\mu, q^\nu] = 2\pi i \alpha' \theta^{\mu\nu}.
\eeq
 Now let us take a look at the Lorentz symmetry of the model. If the
 $B$-field is regarded as a constant background then the Lorentz
 symmetry is broken down to such transformations which preserve
 the form of $B$. To analyze the question we can perform  $SO(d-1,1)$
 transformation which brings $B$ to the canonical block diagonal form
\beq
\label{k4}
 (B_{\mu\nu}) = \left ( \begin{array}{lllllll}
            0 &  \lambda_1 &  0&  0 & ...& 0 & 0 \\
            -\lambda_1 & 0 & 0 & 0& ... & 0 & 0 \\
             0 & 0 & 0 & \lambda_2 & ... & 0 & 0 \\
             0 & 0 & -\lambda_2 & 0 & ... & 0 & 0\\
             ...& ...& ...& ...& ...& ...& ...\\
               0 & 0 & 0 & 0 & ... & 0 & \lambda_{d/2}\\
               0 & 0 & 0 & 0 & ... & -\lambda_{d/2} & 0
             \end{array} \right )
\eeq
 and thus $G_{\mu\nu}$ will have the following form
\beq
\label{k5}
 (G_{\mu\nu}) = \left ( \begin{array}{lllllll}
              \lambda^2_1 - 1& 0 &  0&  0 & ...& 0 & 0 \\
             0 & 1-\lambda_1^2 & 0 & 0& ... & 0 & 0 \\
             0 & 0 & 1+ \lambda^2_2 & 0 & ... & 0 & 0 \\
             0 & 0 & 0& 1+\lambda^2_2  & ... & 0 & 0\\
             ...& ...& ...& ...& ...& ...& ...\\
               0 & 0 & 0 & 0 & ... & 1+\lambda^2_{d/2} & 0\\
               0 & 0 & 0 & 0 & ... & 0 & 1+\lambda^2_{d/2}
             \end{array} \right ) 
\eeq
 where $\eta_{\mu\nu} = (-1, 1, ..., 1)$.
 In other words we rewrite the action (\ref{a1}) in
  coordinates such that $B$ assumes the above form.
 In the Euclidian version of the theory one should replace $G_{00}$ and
 $G_{11}$ by $1+\lambda_1^2$. In the Minkowski case
 $G_{\mu\nu}$ can acquire a $0$ eigenvalue when $\lambda_1 = 1$. 
In this case we will have problems with imposing
 the canonical commutation relations (\ref{b3}) on the model. 
 Thus one can conclude that the model is not well-defined 
 at all values of $\lambda_1$.  

 By looking at (\ref{k4}) one can find the Lorentz group of the model. 
 If $\lambda_1 = \lambda_2 = ... = \lambda_r=0$ and the rest are different
 from zero 
 then the Lorentz group will be $\Lambda= SO(2r-1,1) \otimes
  (SO(2))^{(d/2-r)}$. 
  By definition the new Lorentz group $\Lambda$ will preserve 
 the form of $G_{\mu\nu}$ and $\theta^{\mu\nu}$. 
 Thus any constructions with these
 objects are Lorentz covariant in the present model. The natural
 scalar product of the Lorentz group is given by 
 $k_\mu (G^{-1})^{\mu\nu} p_\nu$
 where $k_\mu$ and $p_\nu$ are vectors which transform under the
 new Lorentz group $\Lambda$. As a result we will see that in mass spectrum
 all particles are classified under the irreducible representations of
 the new Poincar\'e group which is a semidirect product of the new Lorentz
 group $\Lambda$  and the ten dimensional group of translations.      

 To understand better the commutation relations (\ref{k3}) it is
 useful to write down the explicit form of $\theta^{\mu\nu}$ in 
 the basis where $B_{\mu\nu}$ has the form (\ref{k4}) 
\beq
\label{k6}
 (\theta^{\mu\nu}) = \left ( \begin{array}{lllllll}
              0 & \frac{\lambda_1}{1-\lambda^2_1 } &  0 &  0 & ...& 0 & 0 \\
             -\frac{\lambda_1}{1-\lambda_1^2} & 0 & 0 & 0& ... & 0 & 0 \\
             0 & 0 & 0 & -\frac{\lambda_2}{1+ \lambda^2_2}  & ... & 0 & 0 \\
             0 & 0 & \frac{\lambda_2}{1+\lambda^2_2} & 0  & ... & 0 & 0\\
             ...& ...& ...& ...& ...& ...& ...\\
               0 & 0 & 0 & 0 & ... & 0 &
             -\frac{\lambda{d/2}}{1+\lambda^2_{d/2}} \\
               0 & 0 & 0 & 0 & ... & 
             \frac{\lambda_{d/2}}{1+\lambda^2_{d/2}} & 0 
             \end{array} \right ) 
\eeq
 Ignoring the components of $B$ field  along the time direction
 we can see that the components of $|\theta^{\mu\nu}|$ are bounded by $1/2$
 and $|\theta^{\mu\nu}|$ takes the maximal value 
 when B-field is $1$ ($|\lambda|=1$). Thus from (\ref{k3}) 
 we get the following uncertanty relation
\beq
\label{k7}
\Delta q^\mu \Delta q^\nu \geq \frac{1}{2} \pi \alpha'
\eeq 
 at $\lambda=1$. We see that the maximal noncommutative effect appears
 when $B$ is of  order  one and the maximal cells have a size of order 
 $\sqrt{\alpha'}$. Therefore in this situation only one degree of freedom
 can be per Plank cell. The general situation can be summarized as follows:
 as the $B$ field ($\lambda$) varies from $0$ to $\infty$ we are going from
 pure Neumann boundary conditions to pure Dirichlet ones. As we are 
 approaching these limiting situations the noncommutativity disappears.  

\section{Mass spectrum}

 We can now analyze the mass spectrum. The Hamiltonian $L_0$ of the system
 is:
\beq
\label{e1}
L_0 = \alpha' p_\mu (G^{-1})^{\mu\nu} p_\nu    
 + \frac{1}{2}\sum\limits_{n\neq 0} n 
 G_{\mu\nu} \alpha_n^\mu \alpha_{-n}^\nu
\eeq  
 where $a_n^\mu = \sqrt{2\alpha'|n|} \alpha_n^\mu$ (for $n\neq 0$) and 
 $p_\mu$ is defined by (\ref{k1}). Arguments from the previous section
 indicate that the correct Casimir for the new Poincar\'e group is $M^2 = -  
 p_\mu (G^{-1})^{\mu\nu} p_\nu$. 
%Since the anomaly is not modified and 
We assume that $\lambda_1^2 < 1$ (so the metric $G_{\mu\nu}$
does not become degenerate) the quantization can be done in the standard
 way by imposing Virasoro conditions on the physical states:
 $(L_0 -1)|Phys\rangle =0$ and $L_n|Phys\rangle =0$ for $n>0$. Therefore
 the expression for the masses of string states is the following
\beq
\label{e2}
 \alpha' M^2 = \sum\limits_{n=1}^{\infty} n (\alpha_{n}^{+})^\mu
 G_{\mu\nu} (\alpha_{n})^\nu - 1 
\eeq
 where we used $(\alpha_{n}^{+})^\mu = \alpha_{-n}^\mu$. Thus one can
 see that we get the usual spectrum except
 that the states will transform under a non-standard Poincar\'e group. 
% Therefore these states cannot be interpreted anyhow in terms of initial
% full ten dimensional  Poincar\'e group
\footnote{If one tries to 
 define the mass operator as $M^2 = - p_\mu \eta^{\mu\nu} p_\nu$ the
 spectrum of the string turns out to be continuous which is not
one would expect for a sensible string theory.}.

 There is a massless state in the model: 
 $\epsilon_\mu(k)  \alpha_{-1}^\mu|o,k\rangle$. 
Requiring that $L_0 -1$ and $L_1$ annihilate the state implies:
\beq
\label{e3}
 k_\mu (G^{-1})^{\mu\nu} k_\nu =0,\,\,\,\,\,\,\,\,\,\,\,\,\,\,\,\,\,
 \epsilon_\mu(k) (G^{-1})^{\mu\nu} k_\nu =0.   
\eeq
The second condition in (25) should be interpreted in same way
 as it is done in QED when by imposing condition $\partial_\mu A^\mu=0$
 on the Fock space one kills unwanted states.
 It is important to stress that $\epsilon_\mu$ is not a massless
 vector in the normal
 sense. It transforms as a massless vector  under the new Lorentz group 
 $\Lambda$.

The next important question is the following, what is the real gauge
 transformation for this massless gauge boson. Recall that in
 ordinary ($B=0$) open string theory gauge transformations amount
 to a shift of the state by a null state (a physical state which is
  orthogonal to all physical states and therefore of zero norm):
\beq
\label{e4} 
 \epsilon_\mu (k) \alpha^\mu_{-1}|0,k\rangle \,\,\,\rightarrow\,\,\,
 \epsilon_\mu (k) \alpha^\mu_{-1}|0,k\rangle +  
 i k_\mu \gamma(k) \alpha^\mu_{-1}|0,k\rangle 
\eeq
In this case one can show that the shifted state has zero norm on shell,
$k^2=0$.  
 In the next section we will explicitly show that a noncommutative 
gauge transformation in the $B\neq 0$ case results in a shift by
a null state as well. 

\section{Noncommutative gauge transformation of massless state}

 In these two last sections we would like to take a naive (but hopefully
 transparent) look at the noncommutative transformations and its relation
 to the deformation quantization.  

 Let us sketch some important properties of commutative and
 noncommutative gauge transformations which we are going to use. 
For the case of usual abelian gauge transformations we have the
 following invariance
\beq
\label{c1}
A_\mu(x)\,\,\, \rightarrow \,\,\, A_\mu(x) + \partial_\mu \lambda(x)
\eeq
 or in momentum represenation
\beq
\label{c2}
 \epsilon_\mu (k)\,\,\, \rightarrow \,\,\,\epsilon_\mu(k) + i k_\mu 
 \gamma(k)
\eeq
 where we are using the following notation
\beq
\label{c3}
 A_\mu (x) = \frac{1}{(2\pi)^{d/2}} \int d^d k\,e^{ikx} \epsilon_\mu (k),
\,\,\,\,\,\,\,\,\,\,\,\,\,\,\,
\epsilon_\mu (k) =  \frac{1}{(2\pi)^{d/2}}
 \int d^d x\, e^{-ikx} A_\mu(x)
\eeq
  and $\gamma$ is the Fourier transform of $\lambda$. The reality condition
 for $A_\mu(x)$ and $\lambda(x)$ corresponds to the following condition
 $\epsilon^*_\mu(k) = \epsilon_\mu(-k)$, $\gamma^*(k)= \gamma(-k)$. 
 The condition $k_\mu \epsilon^\mu(k) =0$ corresponds to the Lorentz 
 condition for $A^\mu$.

 Noncommutative $U(1)$ gauge transformations act as follows:
\beq
\label{c4}
A_\mu\,\,\, \rightarrow \,\,\, A_\mu + \partial_\mu \lambda + 
i \lambda * A_\mu - i A_\mu * \lambda
\eeq
 where the star product has the following standard definition
\beq
\label{c5}
 f(x) * g(x) = e^{ i\pi\alpha'\theta^{\mu\nu} 
 \frac{\partial}{\partial \xi^\mu} \frac{\partial}{\partial \zeta^\nu}}
 f(x+\xi) g(x+\zeta) |_{\xi=\zeta =0}
\eeq
 In the momentum representation we have
\beq
\label{c6}
 \frac{1}{(2\pi)^{d/2}} \int d^dx\,e^{-ikx} (f(x) * g(x)) =
  \frac{1}{(2\pi)^{d/2}}  \int d^d p\,e^{- i\pi\alpha'\theta^{\mu\nu}
 p_\mu k_\nu} \tilde{f}(p) \tilde{g}(k-p) 
\eeq 
 where 
\beq
\label{c7}
 f(x) = \frac{1}{(2\pi)^{d/2}} \int d^d p\,e^{ipx} \tilde{f}(p),
\,\,\,\,\,\,\,\,\,\,\,\,\,\,\,\,\,\,\,\,\,\,
 g(x) =\frac{1}{(2\pi)^{d/2}} \int d^d l\,e^{ilx} \tilde{g}(l) 
\eeq
 Using (\ref{c4}) and (\ref{c6}) one can get the noncommutative 
 gauge transformation in momentum representation
\beq
\label{c8}
\epsilon_\mu (k)\,\,\,\rightarrow\,\,\,\epsilon_\mu(k) + i k_\mu \gamma(k) 
 + \frac{2}{(2\pi)^{d/2}} \int d^d p\,\gamma(p) \epsilon_{\mu}(k-p)
 \sin (\pi\alpha' \theta^{\nu\rho} p_{\nu} k_{\rho}) 
\eeq
 for the case when $d$ is even (in odd case there are some problems).
It is important to note that in noncommutative gauge transformations
 (\ref{c4}) the gauge parameter is not completly arbitrary (unlike the  
 commutative case). In noncommutative case the field strength is
 not a gauge invariant object and therefore to impose the gauge 
 invariance on the action one should require that the gauge potential
 and gauge parameter fall off at infinity fast enough. In the momentum
 space it means that they have appropriate behavior at zero momentum.
 Also all functions in the noncommutative case should be infinitely 
 differentiable.

 To show that the transformation
(\ref{c8}) is a gauge transformation we should show that the
 following state is null:
\beq
\label{c9}
\int d^d p\,\gamma(p) \epsilon_{\mu}(k-p)
 \sin (\pi\alpha' \theta^{\nu\rho} p_{\nu} k_{\rho}) \alpha_{-1}^\mu 
 |o,k \rangle = f_{\mu}(k)  \alpha_{-1}^\mu  |o,k \rangle
\eeq
 or in other words one has to prove that the following two
 properties hold
\beq
\label{c10}
 f_\mu (k) k^\mu = 0,\,\,\,\,\,\,\,\,\,\,\,\,\,\,\,\,\,\,\,
 f_\mu (k) f^{*\mu} (k) = 0,\,\,\,\,\,\,\,\,\,\,\,\,\,\,\,\,\,\,\,
 \epsilon^{*}_\mu (k) f^\mu(k) = 0
\eeq
 when the properties (\ref{e3}) hold. The properties (\ref{c10}) insure
 that the state (\ref{c9}) is a physical and it is orthogonal to all 
 physical states. 
 Let us show that equations (\ref{c10}) are true. Introducing 
 the notation
\beq
\label{c11}
 A(k) =  f_\mu (k) k^\mu = \int d^d p\,\gamma(p) \epsilon_{\mu}(k-p) k^\mu
 \sin (\pi\alpha' \theta^{\nu\rho} p_{\nu} k_{\rho}) 
\eeq
 one can show that $A^{*}(k) = A(-k)$ using the reality conditions
 for $A_\mu$ and $\lambda$. One can equivalently write (\ref{c11}) as:
\beq
\label{c12}
 A(k) =  f_\mu (k) k^\mu = \int d^d p\,\gamma(p) \epsilon_{\mu}(k-p) p^\mu
 \sin (\pi\alpha' \theta^{\nu\rho} p_{\nu} k_{\rho}) 
\eeq
 where we used $\epsilon_\mu(k-p)\,(k-p)^\mu =0$. From (\ref{c12})
 again using the reality conditions one can show that $A^{*}(k) =
 - A(-k)$ and therefore $A(k)=0$. Since $k_\mu f^\mu(k)=0$ is just
 Lorentz gauge condition we require this condition to be true off-shell
 (i.e. even when $k^2 \neq 0$). 
 In the present proof any subtleties are excluded since all functions
 have nice behavior in general including at zero momentum.

 Now we can prove  that the second property in (\ref{c10}) is true.
 The combination $f_\mu(k) f^{*\mu}(k) = G(k^2)$ is Lorentz invariant 
 and therefore it can depend only on $k^2$. We are only interested
 in the value of $G(k^2)$ when $k^2=0$. Thus it is just a constant which 
 one can calculate by picking some concrete value of $k_\mu$ which
 satifies $k^2=0$. We can do it for $k_\mu = (0,0,...,0)$ and
 find that it is zero. The last condition  conditions in (\ref{c10}) 
 can be proven
 in the same way. Thus 
%using on-shell constraint $k^2=0$ 
 we have proven all conditions in (\ref{c10}).
 
To avoid confusion one should understand that
  properties (36) are not true as  equalities
 on  functions. Since we are dealing
  with a Poincar\'e invariant quantum theory
  one should  look at the correlation
 functions of operators. For instance, the last expresion in
 (36) should be understood as the following correlator
 in coordinate space
\beq
\label{AAAS}
 \langle A^\mu (x) \delta A_\mu(y) \rangle = G((x-y)^2)
\eeq       
 or in momentum space
\beq
\label{AAASS}
 \langle \epsilon^\mu (-k) \delta \epsilon_\mu(k) \rangle = G(k^2)
\eeq
 where by $\delta$ we mean a gauge transformation. It is the fact
 that we are dealing with correlation functions in a Poncar\'e invariant
 vacuum give us such a simple dependence on momentum. 

\section{General gauge transformation and deformation quantization}

 One can try to address the following question: what is the general
 form of a null state which corresponds to gauge transformations. 
 In this section we would like to argue that in general a gauge 
 transformation is related to deformation quantization with
 respect to $\alpha'$. The relation between open string model in $B$ field
 background and the deformation quantization was originaly 
  pointed out in \cite{Schomerus:1999ug}. 

There is only one local solution for conditions (\ref{c10}) which 
 correspond to $k_\mu \gamma(k)$. All other solutions will be nonlocal
 in momentum space. Since we are interested in infinitesimal gauge 
 transformations, a null state should be linear in the gauge field 
 $\epsilon_\mu$ and gauge parameter $\gamma$. 
 Thus the general solution has the form
\beq
\label{r1}
 f_\mu(k) = \int d^dp\, \gamma(p) \epsilon_\mu(k-p) G(p, k-p)
\eeq  
 where $G(p, k-p)$ is a kernel with the following property
\beq
\label{r2}
 G^*(-p, k+p) = G(p, -k-p) = G(-p, k+p) .
\eeq
 The property (\ref{r2}) comes from the reality condition
$f^*_\mu(k) = f_\mu(-k) = - f_\mu(k)$.
 We require two extra properties for the function $G(p, k-p)$ 
\beq
\label{r3}
 G(p, -k-p) = - G(p, k-p),\,\,\,\,\,\,\,\,\,\,\,\,\,\,\,\,\,\,\,
 G(-p, k+p) = - G(p, k-p)
\eeq
 which ensures the condition $k^\mu f_\mu =0$ off-shell (i.e.
 even when $k^2 \neq 0$). 
 As a result of (\ref{r3}) we have $G(p,-p) =0$ and it gives us 
  the two properties in (\ref{c10}) on-shell. 
 It is important to notice that
 the function $G(p, k-p)$ is defined up to gauge field and gauge 
 parameter redefinitions
\beq
\label{r3a}
 f_\mu(k) = \int d^dp\, \gamma(p) s(p) 
 \epsilon_\mu(k-p) s(k-p) \frac{G(p, k-p)}{s(p)s(k-p)}.
\eeq
 These redefinitions should not modify the on-shell physics. In
 other words it shoud not modify the on-shell state
\beq
\label{r3b}
 \epsilon_\mu(k) \alpha_{-1}^\mu|o,k\rangle\,\,\,\,\,\,\,
\rightarrow\,\,\,\,\,\,\,s(k)\epsilon_\mu(k) \alpha_{-1}^\mu|o,k\rangle,
\,\,\,\,\,\,\,\,\,\,\,\,s(0)=1
\eeq 
 The function $s(k)$ has a natural expansion in powers of $\alpha'$.
In the coordinate space these redefinitions correspond to the following
 transformation:
\beq
\label{r3c}
 A_\mu \,\,\,\,\,\,\,\rightarrow\,\,\,\,\,\,\,A_\mu + \alpha' D_1(A_\mu)
 + (\alpha')^2 D_2(A_\mu) + ...
\eeq
 where $D_i$ are differential operators. These redefinitions produce
 the group of automorphisms on the space of gauge fields. Since we would
 like to understand the general form of off-shell gauge transformation 
 it is important to look at the possible field redefinitions which do
 not affect the on-shell physics. 

 $G(p,k-p)$ is a dimensionless scalar function and it can be expanded 
 in powers of $\alpha'$. Lorentz invariance requires that
 it is a function of momenta with all indices contracted using  
either $\theta^{\mu\nu}$ or  $(G^{-1})^{\mu\nu}$.
 In coordinate representation the
 expansion has the form
\beq
\label{r4}
 \delta_\alpha
 A_\mu = \alpha' B_1(\alpha, A_\mu) + (\alpha')^2 B_2(\alpha, A_\mu)
 + ...
\eeq 
 where $B_i$ are bidifferential operators, the general possible form
 can be easily written down. The null states (\ref{r1}) should correspond
 to some gauge symmetry. The finite gauge transformations belong
 to a group. Therefore at level of  infinitesimal transformations   
  one has to  require:
\beq
\label{r4a}
\delta_\alpha \delta_\beta A_\mu - \delta_\beta \delta_\alpha A_\mu=
 \delta_{\{\alpha,\beta\}} A_\mu 
\eeq
 where $\{\alpha,\beta\}$ is the bracket with respect to some associative
 algebra. It turns out that the requirement (\ref{r4a}) is quite 
 restrictive for the form of the kernel $G(p,k-p)$. From (\ref{r4a})
 in the momentum space
 one can read off the bracket as follows
\beq
\label{r4aa}
 \{\alpha,\beta\}(p) = \int d^d\tilde{p}\,\alpha(\tilde{p})
 \beta(p-\tilde{p}) \left[ \frac{G(p-\tilde{p})G(\tilde{p},
 k-\tilde{p}) - G(\tilde{p}, k-p)G(p-\tilde{p}, k-p+\tilde{p})}
 {G(p,k-p)}\right]
\eeq
 The kernel $K(\tilde{p}, p-\tilde{p})$ for the bracket 
 (expression in square brackets) does not depend 
 on the vector $k$ and should be antisymmetric ($K(\tilde{p}, p-\tilde{p}) =
 - K(p-\tilde{p}, \tilde{p})$). Using these requirements one can
 arrive at the conclusion that $K(\tilde{p},p-\tilde{p}) = 
 G(\tilde{p},p-\tilde{p})$. Thus in  coordinate space the gauge
 transformation has the  form of bracket: $\delta_\alpha A_\mu =
 \{\alpha, A_\mu\}$ and the requirement (\ref{r4a}) is just the Jacobi
 identity for this bracket. Using antisymmetry of  $G$ we can 
 write down the leading term of the gauge transformation as
 follows
\beq
\label{r4ab}
 \delta_\alpha A_\mu = \alpha' G_1 \theta^{\mu\nu} \partial_\mu \alpha
 \partial_\nu A_\mu + O(\alpha'^2).
\eeq 
 
 In general a bracket can be defined in terms of some associative product
$*$ as follows:
\beq
\label{r4ac}
\{f_1,f_2\} = f_1*f_2 - f_2*f_1,
\eeq
and let $g$ be the kernel of the product in momentum space:
\beq
\label{r4ad}
 \frac{1}{(2\pi)^{d/2}} \int d^dx\,e^{-ikx} (f_1(x) * f_2(x)) =
  \frac{1}{(2\pi)^{d/2}}  \int d^d p\,e^{- i\pi\alpha'\theta^{\mu\nu}
 p_\mu k_\nu} \tilde{f}_1(p) \tilde{f}_2(k-p)g(p,k-p) 
\eeq 
 where 
\beq
\label{r4ae}
 f_i(x) = \frac{1}{(2\pi)^{d/2}} \int d^d p\,e^{ipx} \tilde{f}_i(p).
\eeq
 The $G$ and $g$ kernels are related in the following simple way
\beq
\label{r4af}
 G(p, k-p) = g(k, k-p) - g(k-p, p)
\eeq
 Thus one can see that by looking for the general gauge transformation
 we are arriving at the problem of classification of associative
 product which is defined as power series in $\alpha'$. We are 
 interested in this product up to field redefinitions (\ref{r3c}).
This is exactly the  star product construction considered 
 by Kontsevich \cite{Kontsevich:1997vb}. Due to Kontsevich in 
 the present case
 the appropriate star product (up to field redefinitions)
 is given by Moyal product.     

 However we can analyze the concrete form of the star product
 in the strightforward way. 
Since $g(p,k-p)$ is a scalar it only depends on scalars constructed from
$k$ and $p$
by contracting indices with either the metric $G_{\mu\nu}$ 
 or the anti-symmetric $\theta^{\mu\nu}$. Thus $g$ is the function
 of four possible kinematic invariants which one can construct from
 vectors $p$ and $k-p$.      
Associativity of the $*$-product implies the following constraint
on the kernel $g$:
\beq
\label{ppp1}
g(q,p-q)g(p,k-p)=g(p-q, k-p)g(q,k-q).
\eeq
 To analyze (\ref{ppp1}) one should rewrite everything in terms of
 all possible kinematic invariants. 
 In turns out that associativity written in this form is strong enough 
 to restrict $g$ to be of the form:
\beq
g(p,k-p) = e^{i C \alpha' p_{\mu}\theta^{\mu\nu}q_{\nu}}F(k^2).
\eeq
 where we have used the field redefinitions in form (\ref{r3a}). 

Expressing $G$ as the kernel of the bracket with respect to 
the $*$ product then gives exactly the non-commutative gauge
transformations of the previous section up to some trivial redefinitions.  
 Thus we have shown
in the other direction that null states corresponding to gauge
transformations are always of the form (\ref{c4}). 

 Certainly the present analysis is naive and to make it precise
 one should show that the corresponding null states have the form 
 $Q_{BRST}|\Psi\rangle$ where $Q_{BRST}$ is the BRST generator. The standard 
 BRST analysis for the open string theory
  is not the right one to address the question
 of noncomutativity. As we saw the noncomutativity involves
 all orders of $\alpha'$. Therefore one should modify somehow the
 standard BRST approach to address these questions. Previously 
 from other reasoning it was argued that standard  
 BRST complex should be modified \cite{RR}. We hope to come back to this 
 question elsewhere.  

%The condition (\ref{r4a}) imposes some restrictions on
% the form of $G(p, k-p)$. Thus the first order in $\alpha'$ has the form
%\beq
%\label{r4b}
% \delta_\alpha  A_\mu = \alpha' G_1 \theta^{\nu\rho}\, \partial_\nu \alpha\, 
% \partial_\rho A_\mu + O(\alpha'^2)
%\eeq  
% which corresponds to the Poisson structure with constant coeffients
% on the flat space. We see that one must have the associative algebra
% with some linear product $*$ which modulo $O(\alpha'^2)$ coincides
% with Poisson structure given by $\theta^{\mu\nu}$. We are interested 
% in this product $*$ up to the field redefinitions (\ref{r3c}).  
% Thus we ended up with star product construction considered 
% in \cite{Kontsevich:1997vb}. Due to Kontsevich in the present case

% the appropriate star product (up to field redefinitions)
%  is given by Moyal product.   

% (comments on the closed string case)

\section{Conclusions and discussion}

In this paper we have analyzed open string theory in the presence
of a constant $B$-field.  We discussed the space-time symmetries
of the problem and discussed the gauge invariance of the massless
vector boson state using naive covariant quantization.  Our main motivation
for doing this was to re-derive the non-commutative gauge invariance of
the system in a transparent way. Following the logic
 of two last sections one can expect as well some noncommutative
 properties for gauge transformations of gravitational and
 antisymmetric fields. This noncommutativity for
 $B$-field was indicated
 recently in \cite{Sheikh-Jabbari:1999ba}.

It is straightforward to analyze closed strings in the presence of
a $B$-field which has a flux through a non-trivial 
2-cycle. From the logic presented in last two sections
  one might expect as well
 some noncommutative properties for the closed strings. 
 This theory has some interesting features and will be discussed
in a subsequent article.

\begin{flushleft} {\Large\bf Acknowledgments} \end{flushleft}

\noindent The research of AF was supported by Swedish National
 Research Council.   

%\newpage

%\vskip 0.5in

\end{document}